# Random terahertz metamaterials

Ranjan Singh,[1] Xinchao Lu,[1] Jianqiang Gu,[1,2] Zhen Tian,[1,2] and Weili Zhang[1,a]

[1]*School of Electrical and Computer Engineering,*
*Oklahoma State University, Stillwater, Oklahoma 74078, USA*
[2]*Center for Terahertz waves and College of Precision Instrument and Optoelectronics*
*Engineering, Tianjin University, and the Key Laboratory of Optoelectronics Information and*
*Technical Science (Ministry of Education), Tianjin 300072, People's Republic of China*

(*September 18, 2009*)

## Abstract

Using terahertz time domain spectroscopy we investigate the normal incidence transmission through periodically and randomly arranged planar split ring resonators (SRRs). Introduction of positional disorder in metamaterials has no effect on the quality factor of the fundamental Inductive-Capacitive (LC) resonance. The dipole resonances undergo broadening and shift in their resonance frequencies. The experiment reveals that the randomly distributed SRR structures interact incoherently at LC resonance but couple coherently at the higher frequency dipole resonance.

[a]Electronic mail: weili.zhang@okstate.edu

## I. INTRODUCTION

Negative index materials have recently attracted a great deal of attention from both experimental and the theoretical aspects due to their sophisticated implications in realizing so called subwavelength focusing and many other unusual wave propagation phenomena *[1,2]*. The vast majority of implementations of negative index materials are based on split ring resonators (SRRs) having electromagnetic properties which cannot be found in the naturally occurring materials *[3]*. SRRs are artificially fabricated and periodically arranged highly resonant structures with sizes much smaller than the wavelengths. During the past decade there has been growing interest in the studies of wave propagation through random media due to their fundamental applications in photonics and biomedical research *[4]*. Some existing metamaterials are anisotropic in nature. However, for most practical applications across the entire electromagnetic spectrum isotropic materials are required and isotropic medium is entirely based on randomly distributed cells in the volume of the host structure. Many of the previous works on metamaterials were focused on investigating the unique properties of wave propagation through periodic structures. There has been a tremendous curiosity to explore the collective behavior of SRRs when a disorder sets in either their geometrical dimensions or periodic positioning on the host media *[5-9]*. An in depth understanding of the influence of randomness on the electromagnetic properties of SRRs can further open the doors towards the development of negative index materials and three dimensional isotropic media in a broad spectrum of electromagnetic waves.

To address the question of how tolerant the metamaterials are to the destruction of their positional periodicity, we investigate the transmission properties of terahertz radiation as it propagates through the disordered SRRs. Recently, the resonances in SRRs at normal incidence has been re-interpreted as the different odd and even eigen modes of the plasmonic excitations of the entire SRR structure depending on the orientation of the incident electric field. The electric field parallel and perpendicular to the SRR gap arm gives rise to the odd and even eigen modes respectively *[10]*. In our experiment we showed for the first time in the terahertz regime that the LC resonance of the SRRs is unaffected by the disordered arrangement. The LC resonance is the lowest order odd eigen mode and it has special importance since for the in-plane illumination this particular mode allows to modify the effective permeability of the metamaterial structure. The next order odd eigen mode and the lowest order even eigen mode, however, are affected by the randomness, causing the resonance dips to decrease in strength.

## II. EXPERIMENTS

We performed measurements using terahertz time-domain spectrometer (THz-TDS) *[11]* with a semi-insulating GaAs photoconductive transmitter and a silicon-on-sapphire (SOS) photoconductive receiver gated by a femtosecond Ti: Sapphire laser with an incident laser beam power of 10 mW on both switches. The system consists of four parabolic mirrors configured in an 8-F confocal geometry that enables a 3.5 mm diameter frequency independent beam waist for small sample characterization *[12]*. As shown in Fig. 1(a)–(d), four sets of planar metamaterial samples with 180 nm thick Al were fabricated by conventional photolithography on a silicon substrate (0.64-mm-thick, p-type resistivity 20 Ω cm) *[13]*. Figures 1(e) and 1(f) shows the schematic of unit cells of single and double SRR with a minimum feature $d = 2$ μm in the splits of the rings and other dimensions $w = 3$ μm, $t = 6$ μm, $l = 36$ μm, and a lattice constant $P = 53$ μm. For the single SRRs, only the outer ring with the same dimensions is retained.

The disorder in arrangement was introduced by manually changing the lattice points of SRRs in a random fashion during the mask layout, ensuring that the SRRs do not touch one another. Each of the periodic SRRs was initially at a lattice position, $\vec{r}_n$, where $\vec{r} = x\vec{i} + y\vec{j}$ and individual SRRs were then displaced by $\pm\vec{\delta x}$ and $\pm\vec{\delta y}$ to introduce positioning disorder as shown in fig. 1a and 1b. Thus, $\vec{\delta r} = \delta x\vec{i} + \delta y\vec{j}$ is used to define the degree of randomness *[8]*. In our case, $|\vec{\delta x}| \leq 32$ μm and $|\vec{\delta y}| \leq 32$ μm, leading to a maximum disorder $|\vec{\delta r}| \leq 45.25$ μm, which is equivalent to 85% of the periodicity, $P = 53$ μm. The random SRRs retain the number density as that of the counterpart periodic SRRs, giving the same metal volume filling fraction, $f = 0.25$ for the single SRRs, *SR* (periodic) and *SR-rd* (random), $f = 0.33$ for the double SRRs, *2SR* (periodic) and *2SR-rd* (random). In the THz-TDS measurements, all the samples were characterized by placing those midway between the transmitter and receiver modules in the far-field at the beam waist and the terahertz wave penetrates the SRRs at normal incidence. The orientation of all the SRR samples was aligned such that the terahertz electric field is first parallel and then perpendicular to the gap-bearing sides of the rings. A bare piece of silicon, identical to the sample substrate, was used as a reference. All the measurements of the transmitted electric field and phase were recorded in time domain.

### III. RESULTS AND DISCUSSION

Figure 2 illustrates the frequency dependent amplitude transmission and the phase spectra for all the periodic and random samples oriented with incident electric field along the SRR gap arm and along non gap arms. The frequency dependent amplitude transmission was extracted from the ratio of the Fourier transformed sample and reference measurements, $|\tilde{t}(\omega)| = |E_{sample}(\omega)|/|E_{reference}(\omega)|$ and the phase was defined as $\arg(\omega) = \angle[E_{sample}(\omega)/E_{reference}(\omega)]$.

The frequency-domain spectrum of *SR*, as shown in Fig. 2(a) reveals mainly two resonant features. The lowest order odd eigenmode or the so called *LC* resonance occurs at 0.54THz with a 20.3% transmission minimum and a resonance line width (FWHM) of 79 GHz. The second order odd eigenmode (dipole) resonance is at 1.56 THz with a 12% transmission and a 117 GHz line width. As shown in Fig. 2(b), the lowest order even eigen mode (dipole) resonates at 1.34 THz.

For the random sample *SR-rd* in the parallel orientation, as shown in Fig. 2(a), it reveals no change in the *LC* resonance feature compared to that of *SR*. The next higher resonance, however, blue shifts by 16 GHz with a line width broadening by 80 GHz and a significant transmission modification from 12% to 28%. In perpendicular orientation the lowest order eigen mode of the random structure, *SR-rd* red shifts by 34 GHz and broadens by 106 GHz. This behavior indicates that the dipole resonance is clearly affected by the positional disorder of SRRs.

Figures 2(c) and 2(d) illustrate the frequency-dependent amplitude transmission of the double SRR samples, *2SR* and *2SR-rd*. Similar results are observed in the periodic and aperiodic double SRRs as that of the single SRRs. The *LC* resonance of *2SR* and *2SR-rd* was found identical in all respects. The second order odd eigen resonance at 1.58THz blue shifted by 16 GHz, broadened by 99 GHz from 0.109 to 0.208 THz, and the resonance minimum decreased by 17.5%. The lowest even eigen mode red shifts by 28.5 GHz and broadens by 49 GHz.

The *LC* resonance is independent of randomness in the positioning of SRRs as its peak frequency and resonance strength remain unaffected. Such resonance arises due to the circular current distribution in the metallic arms of each SRR as revealed in Fig. 3(a). The individual SRRs can be treated as an equivalent *RLC* circuit in which the resistance *R* comes from the resistive metal arms, the inductance *L* arises from the current circulating around the SRR perimeter, and the capacitance *C* is due to the accumulation of charges across the SRR gaps. The amount of current circulating depends on the impedance of the SRR metal arms. The *LC* resonance strength can only be altered if the impedance of the SRR changes *[14-18]*. All the parameters which determine the resistance of the rings remain unchanged for the periodic and their corresponding random counterpart structures, leading to the same current distribution profile in them. The only difference between them being the displaced position of the current loops in the random SRRs. The response of the random SRRs at the *LC* resonance is identical to that of their periodic counterpart as long as their number density remains constant by maintaining equal volume metal filling fraction. The inter SRR interaction between the individual elemental SRR is negligible since the collective behavior of the periodic and the random metamaterials is just a direct addition of individual SRR contributions. This behavior has led to the SRRs being called as *incoherent metamaterials* at the LC resonance mode *[5]*.

The *LC* resonance of SRRs has the form $\omega_{LC} = (c_0/l)[d/(\varepsilon_c t)]^{1/2}$, which is determined primarily by the size of the rings, where, $\varepsilon_c$ is the permittivity of the boarding material across the capacitive gap and $c_0$ is the velocity of light. The geometrical size of the SRRs is not changed while randomizing the structures, which ensures that the inductance and capacitance of SRRs in even the disordered samples remain constant. This explains the reason for not being able to see a shift in the *LC* resonance frequency for the randomly placed structures.

Unlike the *LC* resonance, the dipole resonances in the random structures undergo a change compared to that of the periodic SRRs. This can be expected due to different current profiles the SRRs have at *LC* and the dipole resonance frequencies. The second odd and lowest even eigen mode resonances in the transmission spectra of all the SRRs are due to linear currents in the SRR side arms, shown in Fig. 3(b) and 3(c) , and have similar current density distributions and therefore those resonances undergo similar spectral modification when disorder is introduced in the SRRs. The incident terahertz electric field excites plasmon oscillations of conduction electrons at the surface of individual metallic SRR arms that are parallel to the illuminating field, producing a collection of oscillating dipoles with a dipole moment, $p(t)$. The electric field of an oscillating dipole consists of near field, intermediate field, and far field components and can be expressed as $E(r,t) = \frac{1}{4\pi\varepsilon_0}\left(\frac{p(t)}{r^3} + \frac{1}{r^2 c_0}\frac{\partial p(t)}{\partial t} + \frac{1}{r c_0^2}\frac{\partial^2 p(t)}{\partial t^2}\right)$, with $r$ being the distance from the dipole [19]. The dipole resonance is due to the dipole-dipole interaction between the SRR arms along the direction of the incident E field. The plasmon energy and the resonance line width strongly depend on the inter particle distance. The dipole resonance wavelength in the metamaterial structures becomes comparable to the inter particle distance and as the distance between the dipoles is varied, there is formation of coupled plasmon mode which alternates between superradiative and subradiative behavior modifying the resonance spectral width and causing the blue or red shifting [20,21]. The dipole coupling among the randomized SRRs is highly complicated due to their random positioning with respect to each other and is influenced by number of other factors, hence rendering a straightforward explanation impossible. In the periodic structures, the coupling between the dipoles is stronger compared to the randomly distributed SRRs since the interaction between the oscillating particles is partially cancelled by the disorder, which leads to weaker coupling among the dipoles and higher radiative damping. As a result we observe broadened and weaker plasmon resonances in the randomly scattered SRR metamaterials [22]. The same SRR structure behaves *coherently* at the dipole resonance.

**IV. CONCLUSION**

In conclusion, we have investigated the propagation of electromagnetic waves through disordered metamaterials at terahertz frequencies. We experimentally demonstrate that the spectral location, the line shape and the strength of the *LC* resonance of single and double randomly spaced SRRs remain unaltered when compared to their periodic counterparts having the same volume filling fraction. The circular currents in the split rings add up equally in the periodic and random SRRs so as to give identical collective response stating that the currents in the meta-molecules at *LC* resonance interact incoherently. The radiation losses are observed only at the dipole resonances and they become weaker in the random SRRs due the weak coherent coupling between the randomly scattered oscillating dipoles. The finding here clearly reveals that the random SRRs interact incoherently at the *LC* resonance frequency but coherently at the dipole resonance. Random SRRs would lead to the development of practical three dimensional optically isotropic medium. This experiment in particular demonstrates that the THz metamaterials for various applications can be made based on random distribution of SRR particles instead of traditionally used periodic arrangements.

**ACKNOWLEDGMENTS**

We thank N. I. Zheludev for fruitful discussions. This work was partially supported by the U.S. National Science Foundation, the China Scholarship Council, and the National Basic Research Program of China (Grant Nos. 2007CB310403 and 2007CB310408).

**Figure Captions**

**FIG. 1.** (color online) (a)- (d) Lithographically fabricated single and double periodic (*SR*, *2SR*) and random (*SR-rd*, *2SR-rd*) split ring resonators with 180 nm Al SRRs deposited on n-type silicon substrate. (e)- (f) Schematic of single and double SRR unit with dimensions $w = 3$ μm, $t = 6$ μm, $l = 36$ μm, $l' = 21$ μm, $d = 2$ μm, and periodicity $P = 53$ μm.

**FIG. 2.** (color online) (a) and (b) Measured transmission and phase spectrum of *SR*, *SR-rd* and (c), (d) of *2SR*, *2SR-rd* for parallel and perpendicular polarizations, respectively.

**FIG. 3.** Current profile at (a) *LC* resonance, (b) dipole resonance for parallel polarization and (c) dipole resonance for perpendicular polarization in *SR*.

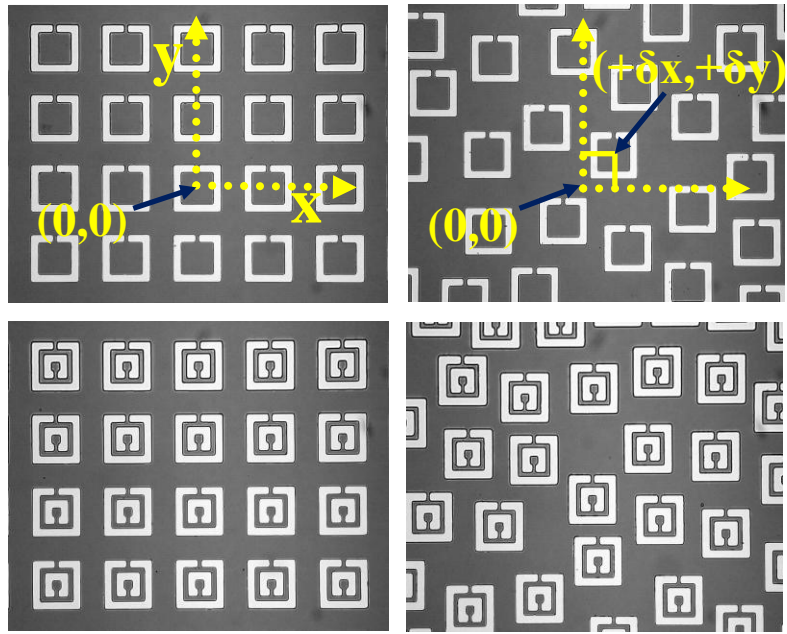

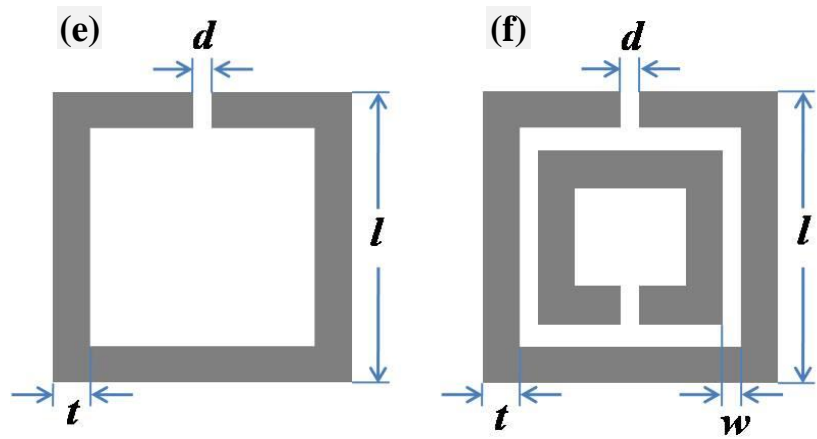

**FIG. 1.**
Singh *et al.*

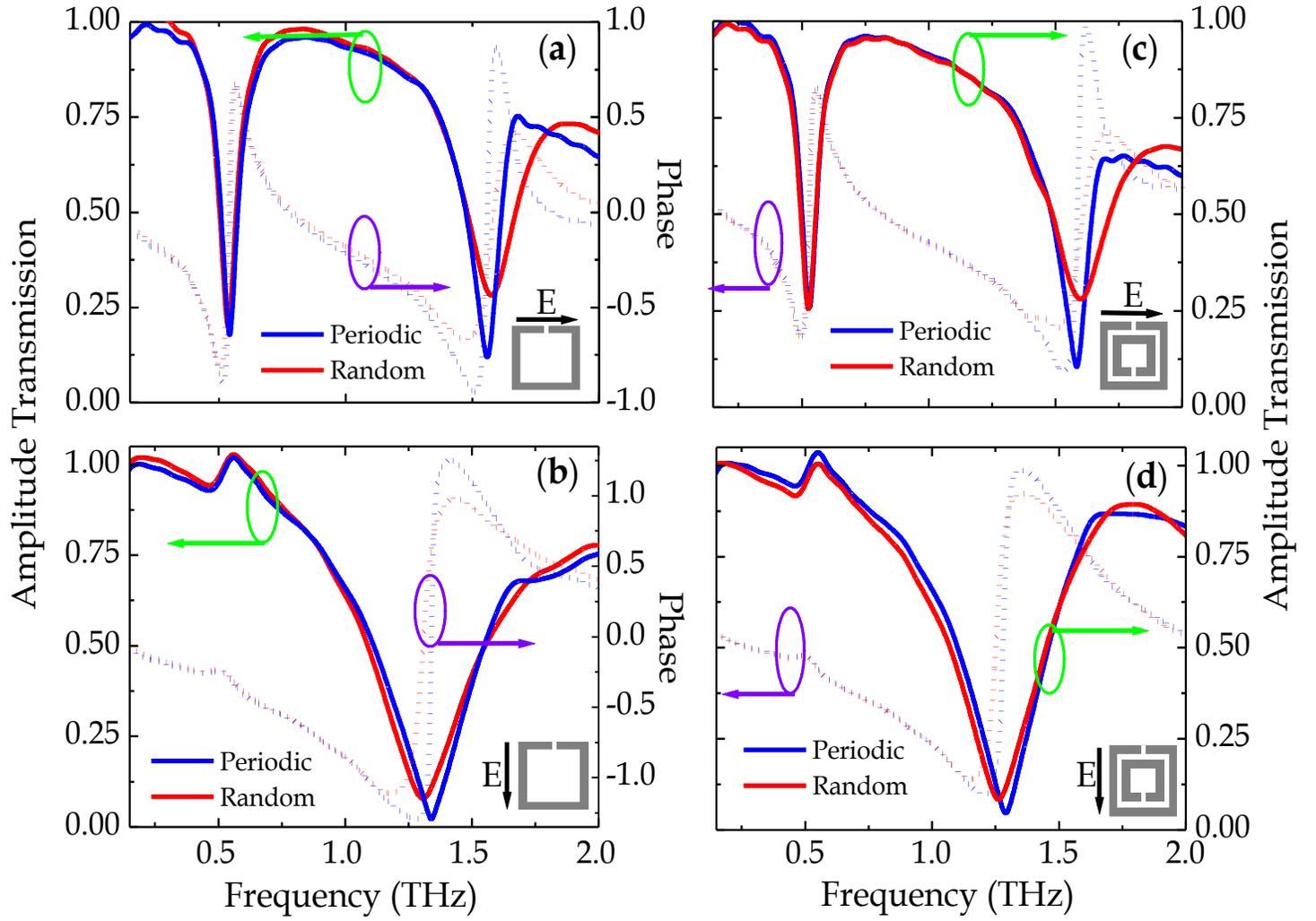

**FIG. 2.**
**Singh *et al.***

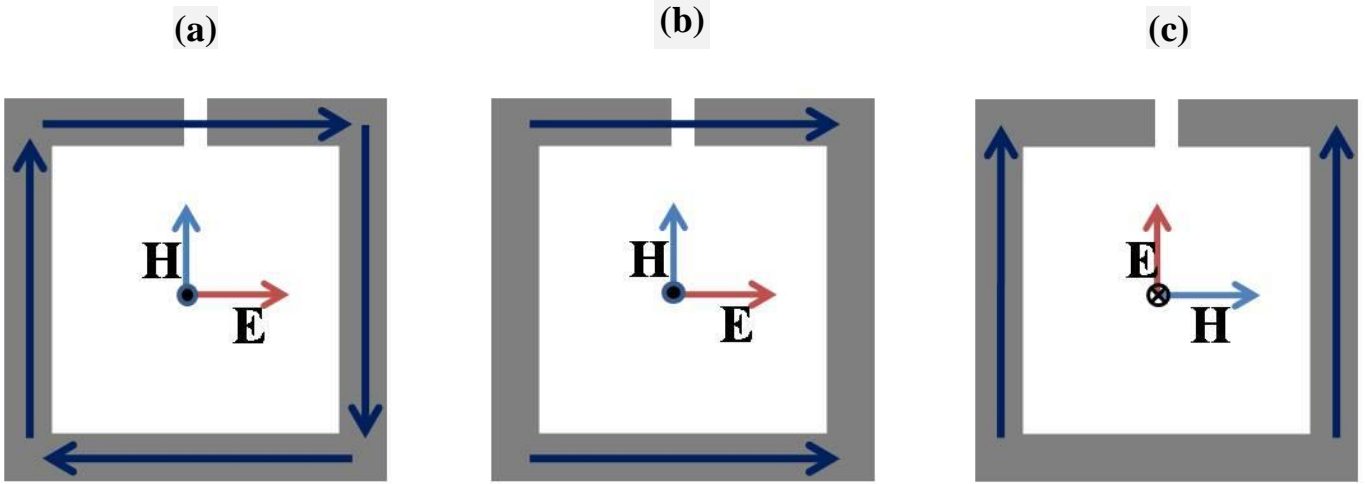

**FIG. 3.**
**Singh** *et al.*